\documentclass[aps,prl,twocolumn,superscriptaddress,showpacs,floatfix,amsmath,amssymb]{revtex4-1}
\usepackage{amssymb}
\usepackage{bm}
\usepackage{url}
\usepackage{graphicx}

\usepackage[dvipsnames,usenames]{color}
\begin{document}
\title{Turbulent fluid acceleration generates clusters of gyrotactic microorganisms} 

\author{Filippo De Lillo}
\affiliation{Dipartimento di Fisica and INFN, Universit\`a di Torino, 
via P. Giuria 1, 10125 Torino, Italy}

\author{Massimo Cencini}
\thanks{Corresponding author}
\email{massimo.cencini@cnr.it}
\affiliation{Istituto dei Sistemi Complessi, Consiglio Nazionale delle 
Ricerche, via dei Taurini 19, 00185 Rome, Italy}

\author{William M. Durham}
\affiliation{Department of Zoology, University of Oxford, 
South Parks Road, Oxford, OX1 3PS, UK}

\author{Michael Barry}
\affiliation{Ralph M. Parsons Laboratory, Department of Civil and 
Environmental Engineering, Massachusetts Institute of Technology, 
77 Massachusetts Avenue, Cambridge, Massachusetts 02139, USA}

\author{Roman Stocker}
\affiliation{Ralph M. Parsons Laboratory, Department of Civil and 
Environmental Engineering, Massachusetts Institute of Technology, 
77 Massachusetts Avenue, Cambridge, Massachusetts 02139, USA}

\author{Eric Climent}
\affiliation{Institut de M\'ecanique des Fluides, Universit\'e de 
Toulouse, INPT-UPS-CNRS, All\'ee du Pr. Camille Soula, F-31400 
Toulouse, France}

\author{Guido Boffetta}
\affiliation{Dipartimento di Fisica and INFN, Universit\`a di Torino, 
via P. Giuria 1, 10125 Torino, Italy}

\begin{abstract}
The motility of microorganisms is often biased by gradients in
physical and chemical properties of their environment, with myriad
implications on their ecology. Here we show that fluid acceleration
reorients gyrotactic plankton, triggering small-scale
clustering. We experimentally demonstrate this phenomenon by studying
the distribution of the phytoplankton {\it Chlamydomonas augustae}
within a rotating tank and find it to be in good agreement with a new,
generalized model of gyrotaxis. When this model is implemented in a
direct numerical simulation of turbulent flow, we find that fluid
acceleration generates multi-fractal plankton clustering, with faster
and more stable cells producing stronger clustering. By producing
accumulations in high-vorticity regions, this process is fundamentally
different from clustering by gravitational acceleration, expanding the
range of mechanisms by which turbulent flows can impact the spatial
distribution of active suspensions.
\end{abstract}


\maketitle 

Microscale patchiness in the distribution of microorganisms has a
profound effect on the ecology of aquatic environments and,
cumulatively, may impact biogeochemical cycling at the global scale
\cite{Azam2007}.  Field observations have revealed that the
centimeter-scale distribution of motile species of phytoplankton is
often considerably more patchy than that of non-motile species
\cite{malkiel1999,gallager2004,mouritsen2003}.  Motility confers
phytoplankton the ability to shuttle between well-lit waters near the
surface during the day and pools of nutrient resources that reside
deeper in the water column at night. This vertical migration is guided
by a stabilizing torque, arising for example from bottom-heaviness,
which tends to keep a cell's swimming direction oriented upwards, and
is contrasted by hydrodynamic shear, which exerts a viscous torque on
cells that tends to overturn them. When the swimming direction results
from the competition between the cell's stabilizing torque and the
shear-induced viscous torque, the organism is said to be gyrotactic
\cite{Pedley1987}.  Gyrotaxis can profoundly affect the spatial
distribution of swimming plankton.  In laminar flows, it produces
remarkable beam-like accumulations in downwelling pipe flows \cite{Kessler1985}
and concentrated layer accumulations in horizontal shear
flows \cite{Durham2009}. In turbulence, gyrotaxis generates intense
microscale clustering at the Kolmogorov scale \cite{Durham2013}.

Previous models of gyrotaxis
\cite{Pedley1987,Hill2002,Durham2009,Thorn2010,Durham2011,Durham2013,Croze2013}
have assumed that the stabilizing torque tends to align the cell
opposite to the direction of gravity.  In intense turbulent flows,
however, fluid acceleration can locally exceed gravitational
acceleration \cite{LaPorta2001}, and turbulence may thus confound the
ability of phytoplankton to ascertain their orientation relative to
the vertical. In this Letter we use a combination of experiments and
modeling to investigate the effect of fluid acceleration on the
distribution of plankton swimming in turbulent flows.

\begin{figure}[t!]
\centering
\includegraphics[width=1\columnwidth]{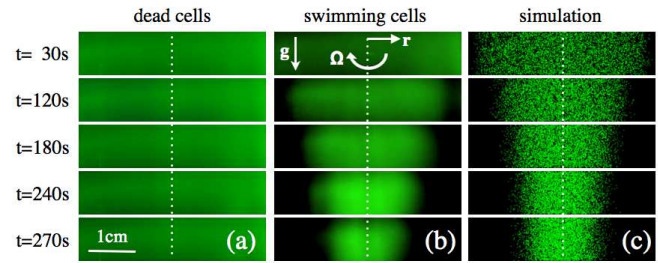}
\caption{(color online) Spatial distribution of gyrotactic swimmers in
  a rotating cylindrical vessel (radius 2 cm, volume 50 ml, rotation
  rate 5 Hz), obtained for (a) cells killed with $8\%$ v/v ethanol,
  (b) swimming cells, and (c) simulated cells. The white dashed line
  denotes the axis of rotation and time is measured since the onset of
  the cylinder's rotation.  (a,b) A culture of {\it C. augustae}
  ($\sim10^5$ cells/ml) illuminated with a green laser (50 mW) sheet.
  Brightness increases with cell concentration.  (c) $10^4$ synthetic
  swimmers, whose positions were obtained by integrating
  Eqs.~(\ref{eq:1})-(\ref{eq:2}) with flow velocity $\bm u=(-\Omega y,
  \Omega x,0)$, with $\Omega=10 \pi \,\mathrm{rad} \, \mathrm{s}^{-1}$
  and cell parameters $v_C=100 \,\mathrm{\mu m} \, \mathrm{s}^{-1}$,
  $B=v_o/g=5 \,\mathrm{s}$ and rotational diffusivity $D_r=0.067
  \,\mathrm{rad}^2\,\mathrm{s}^{-1}$, closely approximating previous
  estimates for \textit{C. augustae} \protect\cite{williams2011}.
\label{fig1}}
\end{figure}

We begin with an illustrative experiment by using a rotating, vertical
cylinder as a simple proxy for a turbulent vortex. The cylinder is
filled with a suspension of \textit{Chlamydomonas augustae}, which, in
a quiescent fluid, migrate upwards against gravity
\cite{williams2011}.  Rotation of the cylinder drives an accumulation
of motile cells at the center of the cylinder, whereas dead cells
remain uniformly distributed (Fig.~\ref{fig1}a,b).  The classic model
of gyrotactic motility \cite{Pedley1987}, which does not include the
effect of fluid acceleration on cell orientation, cannot account for
this simple observation.  A generalized model, which includes the
effect of fluid acceleration, predicts the temporal evolution of the
swimming direction ${\bf p}$ (where $|{\bf p}|=1$) and position $\bm
X$ as
\begin{eqnarray}
\frac{d \bf p}{dt} &=& - \frac{1}{2v_o}\left[\bm A - (\bm A \cdot {\bf p}) {\bf p}\right] 
+\frac{1}{2} {\bm \omega} \times {\bf p}\,
\label{eq:1}\\
\frac{d {\bm X}}{dt} &=& {\bm u} + v_C {\bf p}\,,
\label{eq:2}
\end{eqnarray} 
where ${\bm A}$ is the total acceleration experienced by the cell,
$v_o$ is the characteristic speed with which a perturbed cell
reorients to the direction opposite to ${\bm A}$, ${\bm \omega}={\bm
  \nabla} \times {\bm u}$ is the fluid vorticity at the cell location.
For a bottom-heavy spherical cell $v_o=3 \nu/h$, where $\nu$ is the
kinematic viscosity of the fluid and $h$ the center of mass
displacement from the geometric center.  The cell velocity is the
superposition of the fluid velocity at the cell location, $\bm u$, and
the swimming velocity, $v_C {\bf p}$, where $v_C$ is assumed to be
constant.  We assume that cells are neutrally buoyant, do not impact
the flow, and, owing to their small size ($\sim 10\, \mu m$) can be
modeled as point particles.

In the classic formulation \cite{Kessler1985,Pedley1992} $\bm A = \bm
g=-g\hat{\bf z}$ in Eq.~(\ref{eq:1}) such that a cell's stabilizing
torque aligns motility against gravity.  This model cannot reproduce
the accumulation observed in our experiments, because for solid-body
rotation at angular velocity $\Omega$, one has $\bm \omega= 2\Omega
\hat{\bm z}$ and Eq.~(\ref{eq:1}) predicts that (after a
characteristic orientation time $B=v_o/g$) swimming becomes oriented
along the vertical, ${\bf p} \to \hat{\bm z}$, maintaining the uniform
initial distribution.  Instead, if one accounts for the acceleration
induced by the fluid measured in the reference frame of the particle,
$\bm A=\bm g - \bm a= - g \hat{\bm z}+ \Omega^2 \bm r$, where $(\bm
r,z)$ is the cylindrical coordinate system, the model predicts a
component of cell motility is directed radially inwards.  Indeed,
using the experimental configuration and the known motility parameters
of \textit{C. augustae}, the numerical integration of our model
predicts cell distributions (Fig.~\ref{fig1}c) in close agreement with
those observed (Fig.~\ref{fig1}b), suggesting that our generalization
of the gyrotaxis equations captures the effect of fluid acceleration
on cell motility. The trajectories of cells in Fig.~\ref{fig1}c were
calculated by adding and additional rotational diffusion term
\cite{williams2011} to Eq.~(\ref{eq:1}), which parameterizes the
fluctuations in ${\bf p}$ arising from random cell behavior,
stabilizing cell distribution at finite width about the axis of
rotation at steady state.

The dynamics of this simple experiment, though bearing some
resemblance to persistent small-scale vortices routinely found in
turbulence~\cite{Biferale2005}, cannot capture the complexity of
turbulent flows, which are inherently unsteady and incorporate
multiple scales of fluid motion.  To resolve the role of fluid
acceleration in turbulent flows, we integrate the trajectories of
cells within homogeneous, isotropic turbulence generated via direct
numerical simulations (DNS) of the Navier-Stokes equations
\begin{equation}
\bm a \equiv 
\partial_t {\bm u} + {\bm u} \cdot 
{\bm \nabla} {\bm u} = - {\bm \nabla} p + \nu \nabla^2 {\bm u} + 
{\bm f}\,, 
\label{eq:3}
\end{equation}  
where ${\bm a}$ is the fluid acceleration, $\bm u$ the incompressible
($\bm \nabla\cdot\bm u=0$) fluid velocity, and $p$ the pressure. The
forcing ${\bm f}$ is a zero-mean, temporally uncorrelated Gaussian
random field which injects kinetic energy at large scales at a rate
$\epsilon$, equal to the rate of energy dissipation at small scales
$\epsilon=\nu \langle |{\bm \nabla} {\bm u}|^2\rangle_E$ (where
$\langle[\ldots]\rangle_E= \int d^{3}x [\ldots]$ denotes the Eulerian
average).  We solve Eq.~(\ref{eq:3}) with pseudospectral methods on a
triply periodic cubic domain containing $32^3-256^3$ grid points to
obtain flows with a Taylor Reynolds number of
$Re_\lambda=\sqrt{15}u_{\mathrm{rms}}^{2}/(\nu^{1/2} \epsilon^{1/2})
\approx 20-100$, where $u_{\mathrm{rms}}$ is the root-mean-square
velocity fluctuation.  The Kolmogorov length scale
$\eta_K=(\nu^3/\epsilon)^{1/4}$ of the resulting flow is on the same
order as our grid spacing, ensuring that small-scale fluid motion is
well resolved.

After the flow has reached statistical steady-state, up to $3\times
10^6$ cells with identical $v_C$ and $v_o$ are initialized with random
positions $\bm X$ and orientations ${\bf p}$.  Cell trajectories are
computed by integrating Eqs.~(\ref{eq:1})-(\ref{eq:2}) via
interpolation of fluid velocity, vorticity and acceleration at the
swimmers' position, until cell distributions reach statistical steady
state. Rotational diffusion was not included to reduce the number of
tunable parameters and because the decorrelation timescale due to
stochastic motility ($\sim 15\, s$) is typically longer than
the Kolmogorov timescale of moderately intense turbulence (e.g.
$\tau_K=(\nu/\epsilon)^{1/2}\approx 1\, s$ for $\epsilon = 10^{-6}\, m^2
s^{-3}$).

\begin{figure}[t!]
\includegraphics[width=0.8\columnwidth]{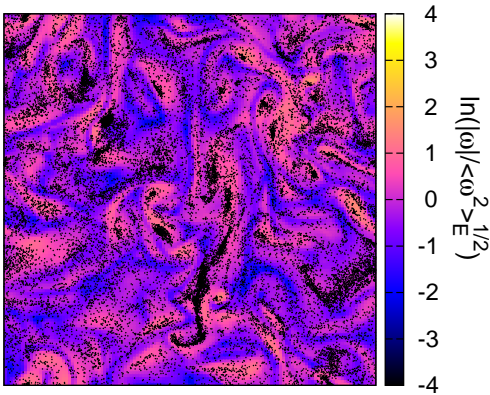}
\vspace{-.4truecm}
\caption{(color online) Slice of a 3D turbulent flow, at
  $Re_\lambda=62$, showing cell clustering (black dots) in high
  vorticity regions when the stabilizing torque aligns them with the
  local fluid acceleration ($\bm A=-\bm{a}$ in Eq.~(\ref{eq:1})), for
  $\Psi_a=1.5$ and $\Phi=1$. Shading shows code the magnitude of the
  fluid vorticity relative to the Eulerian average.}
\label{fig2}
\end{figure}

Two dimensionless parameters characterize cell motility in turbulent
flow. The swimming number $\Phi=v_C/v_{K}$ quantifies the swimming
speed relative to the Kolmogorov velocity $v_K=(\nu
\epsilon)^{1/4}$. The stability number $\Psi_g =
\omega_{\mathrm{rms}}v_o/g$ measures the strength of the viscous
torque exerted by fluid vorticity relative to the stabilizing torque,
where $g$ is taken as the characteristic acceleration scale. While in
general cells are subjected to both gravitational and fluid
acceleration, such that $\bm A=\bm g -\bm a$, we distinguish two
limits. The first limit, $\bm A=\bm g$, considers only the
influence of gravity on cell reorientation: a recent study found that
cells in this regime form clusters in regions of downwelling flow
\cite{Durham2013}.  The second limit $\bm A= -\bm a$ isolates the
effect of fluid acceleration and requires defining a second stability
number (because gravity can no longer be taken as the characteristic
acceleration scale),
$\Psi_a=\omega_{\mathrm{rms}}v_o/a_{\mathrm{rms}}$, where
$a_{\mathrm{rms}}$ is the root-mean-square acceleration
fluctuation. In this limit we find that cells aggregate in regions of
high vorticity (Fig.~\ref{fig2}), revealing that fluid acceleration is
responsible for a second, fundamentally distinct mechanism that drives
clusters of gyrotactic cells in turbulent flow.

Regardless of whether gravitational or fluid acceleration dominates,
the `unmixing' of gyrotactic swimmers by turbulence can be explained
by analyzing the contraction of the cells' phase space, defined by
cell position and swimming
orientation. Equations~(\ref{eq:1})-(\ref{eq:2}) define a dissipative
dynamical system in the $(\bm X,{\bf p})$ phase space of dimension
$2d-1$, and cells inhabit a three dimensional volume such that $d=3$.
One can show the $(\bm X,{\bf p})$ phase space contracts at a rate
\begin{equation}
\Gamma=\sum_{i=1}^{d} \left( \frac{\partial \dot{X}_i}{\partial X_i}+
\frac{\partial \dot{\mathrm{p}}_i}{\partial \mathrm{p}_i} \right)=
\frac{d-1}{2 v_o} \bm A\cdot {\bf p}\,.
\label{eq:4}
\end{equation}
Because the stabilizing torque of gyrotactic swimmers reorients ${\bf
  p}$ towards $-\bm A$, we expect ${\bm A}\cdot {\bf p}$ and,
consequently, $\Gamma$ to be negative on average, indicating that
trajectories will collapse on a fractal attractor in phase space.  If
the fractal dimension of such attractor is less than $d$, its
projection onto the physical space will correspond to clusters with
the same fractal dimension.  A similar phenomenon occurs for inertial
particles, where the contraction of the phase space, defined by
particle position and velocity, leads to fractal clustering
\cite{Bec2005}.  In our case, both a non-zero swimming velocity and a
non-zero stabilizing torque are required for clusters to form, as
$\Gamma \to 0$ for both $\Phi \to 0$ and $\Psi_{g,a} \to \infty$. That
is both non-motile cells and motile cells with no directional bias are
predicted to remain randomly distributed.

\begin{figure}[t!]
\centering
\includegraphics[width=1\columnwidth]{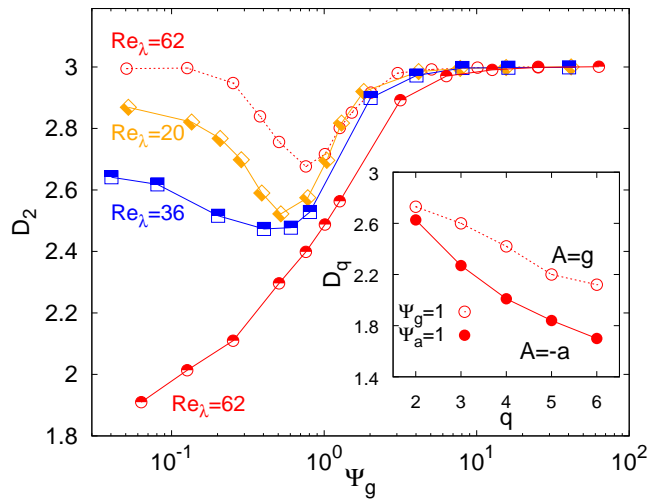}
\vspace{-.6truecm}
\caption{(color online) Correlation dimension $D_2$ versus stability
  number $\Psi_g$ for increasing $Re_\lambda$ (and ratio
  $\alpha=g/a_{\rm{rms}}$) at fixed dimensionless swimming speed
  $\Phi=1/3$.  Semifilled symbols refer to the complete model with
  $\bm A=\bm g-\bm a$ in Eq.~(\ref{eq:1}) with $Re_\lambda=20$
  ($\alpha=0.34$) (orange diamonds), $Re_\lambda=36$ ($\alpha=0.50$)
  (blue squares) and $Re_\lambda=62$ ($\alpha=0.84$) (red circles).
  Empty symbols (red circles) denote the case where cell orientation
  is determined by gravity only, $\bm A=\bm g$ at $Re_\lambda=62$.
  Inset: the generalized dimension $D_q$ as a function of $q$, at
  $Re_\lambda=62$ when only gravitational ($\bm A=\bm g$, empty
  circles) or fluid acceleration ($\bm A=-\bm a$, filled circles) is
  considered.}
\label{fig3}
\end{figure}

To quantify fractal clustering, we measured the correlation dimension,
$D_2$, defined as the scaling exponent of the probability of finding
two cells with a separation distance less than $r$: $P_{2}(|\bm
X_1-\bm X_2|<r)\propto r^{D_2}$ as $r \to 0$
\cite{Paladin1987}. $D_2=d$ denotes randomly distributed cells,
whereas $D_2<d$ indicates fractal patchiness, with smaller $D_2$
corresponding to more clustered distributions and increased
probability of finding pairs of swimmers at close separation.
Figure~\ref{fig3} shows $D_2$ as a function of $\Psi_g$ at different
$Re_\lambda$.  When compared with the case in which the local fluid
acceleration is neglected (${\bf A}={\bf g}$ in Eq.~(\ref{eq:1});
empty symbols in Fig.~\ref{fig3}), these results demonstrate that
fluid acceleration enhances clustering (smaller $D_2$).  These
findings are further supported by measurements of the generalized
fractal dimension $D_q$, which quantifies  the scaling behavior of the
probability of finding $q$ particles within a small separation $r$
\cite{Paladin1987}.  The non-trivial dependence on $q$, $D_q\neq D_2$,
observed in Fig.~\ref{fig3} (inset) indicates that the dynamical
attractor is multifractal~\cite{Paladin1987}.

To formalize the relative contributions of fluid and gravitational
acceleration, it is useful to recast our simulations with different
$Re_\lambda$ in terms of the ratio $\alpha = a_{\mathrm{rms}}/g$.  We
start by briefly summarizing the case $\alpha=0$, when fluid
acceleration is negligible and $\bm A=\bm g$, analyzed in
Ref.~\cite{Durham2013}. In this limit, $D_2$ is insensitive to
$Re_\lambda$ and reaches a minimum (denoting maximal clustering) at
stability numbers, $\Psi_g=O(1)$ (Fig.~\ref{fig3}, open circles),
intermediate between strictly upward motility $(\Psi_g \ll 1)$ and
isotropic motility $(\Psi_g \gg 1)$.  Moreover, one can theoretically
predict that cells preferentially concentrate in downwelling regions
(i.e. where $u_z<0$). Assuming $\Psi_g \ll 1$ allows Eq.~(\ref{eq:1})
to be expanded to first order in $\Psi_g$, obtaining that cells behave
as tracers advected by a velocity field, $\dot{\bm X}={\bm v}({\bm
  X},t)$, that is weakly compressible. Indeed one can show that $\bm
\nabla \cdot \bm v=-\Phi \Psi_g \nabla^2 u_z$.  Cells preferentially
accumulate where $\bm \nabla \cdot \bm v <0$ implying $\nabla^2
u_z>0$, which corresponds to local downwelling flow $u_z<0$, because
$\langle u_z \nabla^2 u_z\rangle_E=-\epsilon/(3\nu)<0$ in isotropic
turbulence (see \cite{Durham2013} for details).  This argument also
correctly predicts that compressibility increases, enhancing
clustering, with the swimming speed $\Phi$ (for small $\Psi_g$) and
vanishes at $\Psi_g=0$.  At large $\Psi_g$, vorticity overturning
dominates and cells swim in random orientations. The balance between
these two mechanisms explain the minimum in $D_2$.

As $\alpha$ increases from zero, the minimum $D_2$ becomes
progressively smaller, indicating more intense clustering, and shifts
towards smaller values of $\Psi_g$, eventually disappearing as
$\alpha$ increases further (Fig.~\ref{fig3}, semifilled
symbols). These results indicate that fluid acceleration substantially
enhances cell clustering, for $\Psi_g\ll 1$ and this effect increases
with the turbulence intensity (larger $Re_\lambda$).

\begin{figure}[t!]
\centering
\includegraphics[width=1\columnwidth]{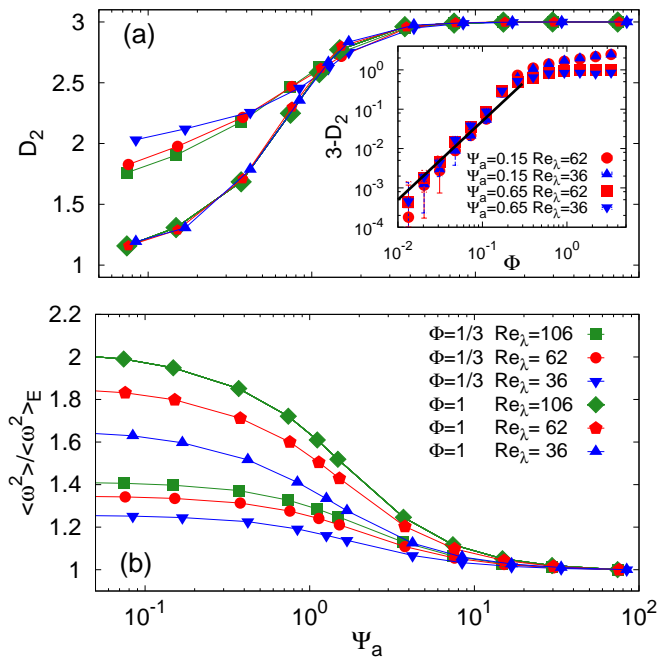}
\vspace{-.5truecm}
\caption{Results of simulations for the model with fluid acceleration
  only, $\bm A=-\bm a$.  (a) Correlation dimension and (b) square
  vorticity averaged over particle positions, $\langle \omega^2
  \rangle$, and normalized by the Eulerian value, $\langle \omega^2
  \rangle_E$, as a function of the stability number $\Psi_a$ for
  different values of $Re_\lambda$ and non-dimensional swimming speeds
  $\Phi$.  Inset of panel (a): the co-dimension $3-D_2$ as a function
  of $\Phi$ at different $Re_\lambda$ and $\Psi_a$ in log-log plot.
  The straight line shows the theoretical prediction $3-D_2 \simeq
  \Phi^2$.
\label{fig4}}
\end{figure}

To understand how fluid acceleration drives clustering, we performed
simulations where ${\bm A}=-{\bm a}$.  In this limit, results for
different $Re_\lambda$ collapse when plotted as a function of $\Psi_a$
(Fig.~\ref{fig4}a), confirming that cells' stability toward fluid
acceleration is the key flow parameter controlling clustering.  Once
plotted as a function of $\Psi_a$, the correlation dimension has only
a weak residual dependence on $Re_\lambda$. In addition, cells cluster
more strongly as cell stability ($1/\Psi_a$) and swimming speed
($\Phi$) increase (Fig.~\ref{fig4}a), corroborating our findings with
the full model (Fig.~\ref{fig3}).  To rationalize these observations
with a theoretical model, we assume $\Psi_a\ll 1$ such that a cell's
stabilizing torque dominates the torque arising from fluid vorticity.
In this limit ${\bf p}$ instantaneously aligns with the local
direction of the fluid acceleration, $\hat{\bm a}={\bm a}/a$, so that
cells move with velocity ${\bm v}\approx {\bm u}+\Phi \hat{\bm a}$,
which is valid to the first order in $\Psi_a$.  While the fluid
velocity $\bm u$ is incompressible, $\bm v$ is not, because ${\bm
  \nabla} \cdot {\bm v} \approx \Phi {\bm \nabla} \cdot \hat{\bm a}
\neq 0$.  Moreover, DNS data show that the sign of $\bm \nabla \cdot
\bm a$ is strongly correlated to that of $\bm \nabla \cdot \hat{\bm
  a}$.  Therefore, when ${\bm A}=-{\bm a}$, gyrotactic cells are
expected to accumulate in regions where $\bm \nabla \cdot \bm a<0$, a
scenario reminiscent of the clustering of nonmotile, buoyant inertial
particles \cite{Balkovsky2001,Calzavarini2008}. These regions
correspond to zones of high fluid vorticity because taking the
divergence of Eq. (\ref{eq:3}) yields $\bm \nabla \cdot \bm
a=\sum_{ij} (\hat{S}_{ij}^2-\hat{\Omega}_{ij}^2)$, with
$\hat{S}_{ij}=(\partial_j u_i+ \partial_i u_j)/2$ and
$\hat{\Omega}_{ij}=(\partial_j u_i-\partial_i u_j)/2$ being the rate
of strain tensor and vorticity tensor, respectively.  The accumulation
in high vorticity regions is demonstrated qualitatively in
Fig.~\ref{fig2} and is quantified in Fig.~\ref{fig4}b, which shows
that the square vorticity averaged over all cell positions is
considerably enhanced over the fluid background value, and increases
with both ${\rm Re}_\lambda$ and $\Phi$. General dynamical systems
considerations~\cite{Falkovich2002,Fouxon2012} predict that, in weakly
compressible flows, the codimension $3-D_2$ has a square power
dependence on the intensity of the divergence of the velocity field.
Therefore, for stable cells ($\Psi_a\ll 1$), in the limit of small
$\Phi$, we should expect $3-D_2 \propto \Phi^2$ as confirmed by our
DNS data (Fig.~\ref{fig4}a inset).

Our results indicate that distribution of gyrotactic swimmers becomes
significantly more clustered when fluid acceleration is on the same
order as gravitational acceleration. However turbulence in natural
environments is often too weak to reach this regime.  For example, the
energy dissipation rate in the ocean rarely exceeds $\epsilon \sim
10^{-4} \, m^2/s^3$, corresponding to $a_{\mathrm{rms}} \simeq
(\epsilon^3/\nu)^{1/4} \simeq 0.03\, m/s^2 \ll g$.  Thus, under most
marine conditions we expect cell distributions can be well
characterized assuming reorientation occurs due to gravitational
acceleration alone, $\bm A= \bm g$ \cite{Durham2013}. We note,
however, that non-homogeneous conditions, such as solid boundaries,
can generate intense vorticity at moderate Reynolds numbers and thus
may drive fluid acceleration induced cell clustering in the bottom
boundary layer. A similar phenomenon may occur in laboratory studies
of plankton, which often employ turbulent dissipation rates much
higher than found in the ocean's upper mixed layer \cite{Peters1997}.
Another prominent environment with intense turbulence occurs in
engineered biofuel production facilities, where turbulent mixing is
used to prevent self-shading and biofouling \cite{Chisti2007}. The
clustering mechanism demonstrated here likely dramatically increases
cell-cell encounter rates and therefore may lead to undesirable cell
aggregates that enhances sedimentation. We finally remark that the
effective compressibility generated by gyrotactic motility in
turbulence may have far reaching implications for population dynamics
and genetics \cite{pigolotti2012,benzi2012} of these tiny inhabitants
of the oceans.

\begin{acknowledgments}
 We thank F. Di Cunto and S. Gallian for help with the experiment,
 M. A. Bees for useful suggestions, and the KITPC institute for
 hospitality during the \textit{New Directions in Turbulence} program
 (to GB and MC). We acknowledge support by MIUR PRIN-2009PYYZM5 and by
 COST Action MP0806 (to GB, FD and MC), by the Human Frontier Science
 Program (to WMD), by the MIT MISTI-France program (to EC and RS), and
 by NSF through grants OCE-0744641-CAREER and CBET-1066566 (to
 RS). \textit{C. augustae} were provided by CCALA, Institute of Botany
 of the AS CR, T\v{e}bo\v{n} Czech Republic.
\end{acknowledgments}


%

\end{document}